\documentclass{modified_aipproc}
\usepackage{amssymb}
\layoutstyle{8x11double}
 
\newcommand{\be}{\begin{equation}}
\newcommand{\ee}{\end{equation}}
\newcommand{\bea}{\begin{eqnarray}}
\newcommand{\eea}{\end{eqnarray}}

\newcommand{\lsim}{
\mathrel{\hbox{\rlap{\hbox{\lower4pt\hbox{$\sim$}}}\hbox{$<$}}}}
\newcommand{\gsim}{
\mathrel{\hbox{\rlap{\hbox{\lower4pt\hbox{$\sim$}}}\hbox{$>$}}}}
 
\def\e6{$E(6)$}
\def\10{$SO(10)$}
\def\21{$SU(2) \otimes U(1) $}

\def\422{$SU(4) \otimes SU(2) \otimes SU(2)$}
\def\321{$SU(3) \otimes SU(2) \otimes U(1)$}
\def\lsim{\raise0.3ex\hbox{$\;<$\kern-0.75em\raise-1.1ex\hbox{$\sim\;$}}}
\def\gsim{\raise0.3ex\hbox{$\;>$\kern-0.75em\raise-1.1ex\hbox{$\sim\;$}}}

\newcommand{\ed}{\end{document}}
\DeclareMathAlphabet{\mathsc}{OT1}{cmr}{m}{sc}

\newcommand{\CL}   {C.L.}
\newcommand{\dof}  {d.o.f.}

\newcommand{\eVq}  {\rm{eV}^2}
\newcommand{\Sol}  {\mathsc{sol}}
\newcommand{\Atm}  {\mathsc{atm}}

\newcommand{\Dms}  {\Delta m^2_\Sol}
\newcommand{\Dma}  {\Delta m^2_\Atm}

\def \nbb {$\beta\beta_{0\nu}$ }

\newcommand{\AddrAHEP}{%
 AHEP Group, Instituto de F\'{\i}sica Corpuscular,
  C.S.I.C. -- Universitat de Val{\`e}ncia \\
  Edificio de Institutos de Paterna, Apartado 22085,
  E--46071 Val{\`e}ncia, Spain\\}

\begin{document}

\title{Neutrino masses and oscillations}

\author{J. W. F. Valle}{
 address={\AddrAHEP} 
}

\begin{abstract}
  
  I summarize the status of three--neutrino oscillations that follow
  from combining the relevant world's data. The discussion includes
  the small parameters $\alpha \equiv \Dms/\Dma$ and
  $\sin^2\theta_{13}$, which characterize the strength of CP violation
  in neutrino oscillations, the impact of oscillation data on the
  prospects for probing the absolute scale of neutrino mass in \nbb
  and the robustness of the neutrino oscillation interpretation itself
  in the presence of non-standard physics. I also comment on the
  theoretical origin of neutrino mass, mentioning recent attemps to
  explain current oscillation data.

\end{abstract}

\maketitle


\section{Introduction}

The discovery of neutrino oscillations has marked a turning point in
our understanding of nature and has brought neutrino physics to the
center of attention of the particle, nuclear and astrophysics
communities.
Here I summarize the determination of neutrino mass and mixing
parameters in neutrino oscillation studies following
Ref.~\cite{Maltoni:2004ei} to which the reader is referred for details
on data analysis and experimental references. For future neutrino
oscillation projects see Ref.~\cite{industry}.
The structure of the three-flavour lepton mixing matrix in various
gauge theories of neutrino mass was given in~\cite{schechter:1980gr}.
Current neutrino oscillation data are well described by its simplest
unitary form, with is no sensitivity to CP violation.  The effect of
Dirac CP phases in oscillations and Majorana phases in \nbb constitute
the main challenge for the future.
The interpretation of the data requires good calculations of solar and
atmospheric neutrino fluxes~\cite{Bahcall:2004fg,Honda:2004yz},
neutrino cross sections and experimental response functions, as well
the inclusion of matter
effects~\cite{mikheev:1985gs,wolfenstein:1978ue} in the Sun and the
Earth.

\section{Solar and KamLAND data}
\label{sec:solar-+-kamland}

The solar neutrino data includes the rates of the chlorine experiment
($2.56 \pm 0.16 \pm 0.16$~SNU), the results of the gallium experiments
SAGE ($66.9~^{+3.9}_{-3.8}~^{+3.6}_{-3.2}$~SNU) and GALLEX/GNO ($69.3
\pm 4.1 \pm 3.6$~SNU), as well as the 1496--day Super-K data (44 bins:
8 energy bins, 6 of which are further divided into 7 zenith angle
bins). The SNO data include the data from the salt phase in the form
of the neutral current (NC), charged current (CC) and elastic
scattering (ES) fluxes, the 2002 spectral day/night data (17 energy
bins for each day and night period) and the 391--day data.  The
analysis includes not only the statistical errors, but also systematic
uncertainties such as those of the eight solar neutrino fluxes.

KamLAND detects reactor anti-neutrinos at the Kamiokande site by the
process $\bar\nu_e + p \to e^+ + n$, where the delayed coincidence of
the prompt energy from the positron and a characteristic gamma from
the neutron capture allows an efficient reduction of backgrounds.
Most of the incident $\bar{\nu}_e$'s come from nuclear plants at
distances of $80-350$ km from the detector, far enough to probe large
mixing angle (LMA) oscillations.
To avoid large uncertainties associated with the geo-neutrino flux an
energy cut at 2.6~MeV prompt energy is applied for the oscillation
analysis.

The first KamLAND data correspond to a 162 ton-year exposure gave 54
anti-neutrino events in the final sample, after all cuts, while $86.8
\pm 5.6$ events are predicted for no oscillations with $0.95\pm 0.99$
background events, consistent with the no--disappearance hypothesis at
less than 0.05\% probability.  This gave the first evidence for the
disappearance of reactor neutrinos before reaching the detector, and
thus the first terrestrial confirmation of oscillations with $\Dms$.
Additional KamLAND data with a somewhat larger fiducial volume of the
detector were presented at Neutrino 2004, corresponding to an
766.3~ton-year exposure.  In total 258 events have been observed,
versus $356.2\pm 23.7$ reactor neutrino events expected in the case of
no disappearance and $7.5\pm 1.3$ background events. This leads to a
confidence level of 99.995\% for $\bar\nu_e$ disappearance.  Moreover
evidence for spectral distortion consistent with oscillations is
obtained.

A very convenient way to bin the latest KamLAND data is in terms of
$1/E_\mathrm{pr}$, rather than the traditional bins of equal size in
$E_\mathrm{pr}$. Various systematic errors associated to the neutrino
fluxes, backgrounds, reactor fuel composition and individual reactor
powers, small matter effects, and improved $\bar{\nu}_e$ flux
parameterization are included~\cite{Maltoni:2004ei}.  This singles out
the LMA solution from the previous ``zoo'' of
alternatives~\cite{Maltoni:2003da}.  The stronger evidence for
spectral distortion in these data also leads to improved $\Dms$
determination, substantially reducing the allowed region of
oscillation parameters.  From this point of view KamLAND has played a
key role in the resolution of the solar neutrino problem.
Assuming CPT invariance one can directly compare the information
obtained from solar neutrino experiments with the KamLAND reactor
results.

\section{Atmospheric and K2K data}
\label{sec:atmospheric-+-k2k}
                                                                               
The first evidence for neutrino oscillations was the zenith angle
dependence of the $\mu$-like atmospheric neutrino data from the
Super-K experiment in 1998, an effect also seen in other atmospheric
neutrino experiments.  However, though appealing, the original
oscillation interpretation was certainly not
unique~\cite{Gonzalez-Garcia:1998hj}. Today, thanks to the
accumulation of upgoing muon data, and the observation of the dip in
the $L/E$ distribution of the atmospheric $\nu_\mu$ survival
probability, the signature for atmospheric neutrino oscillations has
become clear.  The data include Super-K charged-current atmospheric
neutrino events, with the $e$-like and $\mu$-like data samples of sub-
and multi-GeV contained events grouped into 10 zenith-angle bins, with
5 angular bins of stopping muons and 10 through-going bins of up-going
muons.  We do not use $\nu_\tau$ appearance, multi-ring $\mu$ and
neutral-current events, since an efficient Monte-Carlo simulation of
these data would require further details of the Super-K experiment, in
particular of the way the neutral-current signal is extracted from the
data. We employ the latest three--dimensional atmospheric neutrino
fluxes given in ~\cite{Honda:2004yz}.

$\nu_\mu$ disappearance over a long-baseline probing the same $\Delta
m^2$ region relevant for atmospheric neutrinos is now available from
the KEK to Kamioka (K2K) neutrino oscillation experiment.
Neutrinos produced by a 12~GeV proton beam from the KEK proton
synchrotron consist of 98\% muon neutrinos with a mean energy of
1.3~GeV. The beam is controlled by a near detector 300~m away from the
proton target.  Comparing these near detector data with the $\nu_\mu$
content of the beam observed by the Super-K detector at a distance of
250~km gives information on neutrino oscillations.

The data K2K-I sample ($4.8\times 10^{19}$ protons on target) gave 56
events in Super-K, whereas $80.1^{+6.2}_{-5.4}$ were expected for no
oscillations. The K2K-II data correspond to $4.1\times 10^{19}$
protons on target, comparable to the K2K-I sample.  Altogether they
give 108 events in Super-K, to be compared with
$150.9^{+11.6}_{-10.0}$ expected for no oscillations. Out of the 108
events 56 are so-called single-ring muon events.  This data sample
contains mainly muon events from the quasi-elastic scattering $\nu_\mu
+ p \to \mu + n$, and the reconstructed energy is closely related to
the true neutrino energy.  The K2K collaboration finds that the
observed spectrum is consistent with the one expected for no
oscillation only at a probability of 0.11\%, whereas the best fit
oscillation hypothesis spectrum has a probability of 52\%.
                                                                               
One finds that the neutrino mass-squared difference inferred from the
$\nu_\mu$ disappearance in K2K agrees with atmospheric neutrino
results, providing the first confirmation of oscillations with $\Dma$
with accelerator neutrinos. Unfortunately in the current data sample
K2K gives a rather weak constraint on the mixing angle, due to low
statistics.
However, although the determination of $\sin^2\theta_\Atm$ is
completely dominated by atmospheric data, K2K data already start
constraining the allowed $\Dma$ region~\cite{Maltoni:2004ei}.
In particular, there is a constraint on $\Dma$ from below, which is
important for future long-baseline experiments, since these are
drastically affected if $\Dma$ lies in the lower part of the 3$\sigma$
range indicated by current atmospheric data alone.
                                                                               
\section{Three-neutrino oscillations}

The first systematic study of the effective lepton mixing matrix in
gauge theories of massive neutrinos was given
in~\cite{schechter:1980gr}.  For some models this matrix can be taken
as approximately unitary. For three neutrinos, this gives
\begin{equation}
  \label{eq:2227}
K =  \omega_{23} \omega_{13} \omega_{12}
\end{equation}
where each factor is effectively $2\times 2$ and contains an angle and
a CP phase. Two of the three angles are involved in solar and
atmospheric oscillations, so we set $\theta_{12} \equiv \theta_\Sol$
and $\theta_{23} \equiv \theta_\Atm$.  The last angle in the
three--neutrino leptonic mixing matrix is $\theta_{13}$, 
$$\omega_{13} = \left(\begin{array}{ccccc}
c_{13} & 0 & e^{i \phi_{13}} s_{13} \\
0 & 1 & 0 \\
-e^{-i \phi_{13}} s_{13} & 0 & c_{13}
\end{array}\right)\,.
$$
for which only an upper bound currently exists.  All three phases
are physical~\cite{schechter:1981gk}, one corresponds to the one
present in the quark sector (Dirac-phase) and affects neutrino
oscillations, while the other two are associated to the Majorana
nature of neutrinos and show up in neutrinoless double beta decay and
other lepton-number violating processes, but not in conventional
neutrino oscillations~\cite{schechter:1981gk,doi:1981yb}.
                                                                               
Current neutrino oscillation experiments are insensitive to CP
violation, thus we neglect all phases. In this approximation
three-neutrino oscillations depend on the three mixing parameters
$\sin^2\theta_{12}, \sin^2\theta_{23}, \sin^2\theta_{13}$ and on the
two mass-squared differences $\Dms \equiv \Delta m^2_{21} \equiv m^2_2
- m^2_1$ and $\Dma \equiv \Delta m^2_{31} \equiv m^2_3 - m^2_1$
characterizing solar and atmospheric neutrinos.  The hierarchy $\Dms
\ll \Dma$ implies that one can set, to a good approximation, $\Dms =
0$ in the analysis of atmospheric and K2K data, and $\Dma$ to infinity
in the analysis of solar and KamLAND data.
Apart from the data already mentioned, the global oscillation analysis
also includes the constraints from the CHOOZ and Palo Verde reactor
experiments.
                                                                               
\begin{figure}[t] \centering
    \includegraphics[width=.95\linewidth]{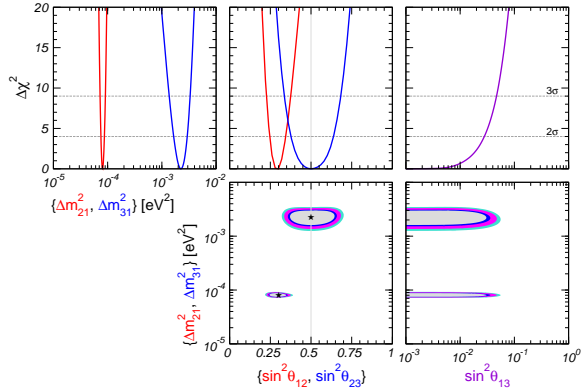}
    \caption{\label{fig:global} %
      Three--neutrino regions allowed by the world's oscillation data
      at 90\%, 95\%, 99\%, and 3$\sigma$ \CL\ for 2 \dof\. In top
      panels $\Delta \chi^2$ is minimized wrt undisplayed parameters.}
\end{figure}

The results of the global three--neutrino analysis are summarized in
Fig.~\ref{fig:global} and in Tab.~\ref{tab:summary}, taken from
Ref.~\cite{Maltoni:2004ei}. In the upper panels of the figure the
$\Delta \chi^2$ is shown as a function of the parameters
$\sin^2\theta_{12}, \sin^2\theta_{23}, \sin^2\theta_{13}, \Delta
m^2_{21}, \Delta m^2_{31}$, minimized with respect to the undisplayed
parameters. The lower panels show two-dimensional projections of the
allowed regions in the five-dimensional parameter space. The best fit
values and the allowed 3$\sigma$ ranges of the oscillation parameters
from the global data are summarized in Tab.~\ref{tab:summary}.  This
table gives the current status of neutrino oscillation parameters.
\begin{table}[t] \centering    \catcode`?=\active \def?{\hphantom{0}}
      \begin{tabular}{|l|c|c|}        \hline        parameter & best
      fit & 3$\sigma$ range         \\  \hline\hline        $\Delta
      m^2_{21}\: [10^{-5}~\eVq]$        & 7.9?? & 7.1--8.9 \\
      $\Delta m^2_{31}\: [10^{-3}~\eVq]$        & 2.2?? &  1.4--3.3 \\
      $\sin^2\theta_{12}$        & 0.31? & 0.24--0.40 \\
      $\sin^2\theta_{23}$        & 0.50? & 0.34--0.68 \\
      $\sin^2\theta_{13}$        & 0.000 & $\leq$ 0.047 \\
      \hline
\end{tabular}    \vspace{2mm}
\caption{\label{tab:summary} Current oscillation parameters.}
\end{table}

As it has long been noted, in a three--neutrino scheme CP violation
disappears when two neutrinos become
degenerate~\cite{schechter:1980gr} or when one angle vanihes, such as
$\theta_{13}$~\cite{schechter:1980bn}.  All genuine three--flavour
effects involve the mass hierarchy parameter $\alpha \equiv \Dms/\Dma$
and the mixing angle $\theta_{13}$.
\begin{figure}[t]
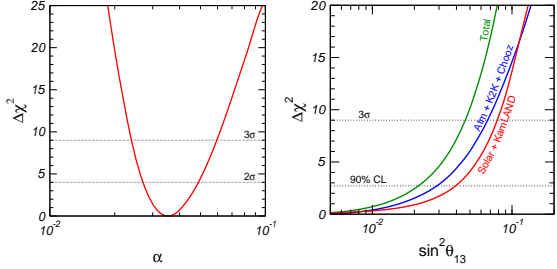
 \centering
\includegraphics[height=3.5cm,width=.45\linewidth]{F-fcn.alpha-2.eps}
\includegraphics[height=3.5cm,width=.45\linewidth]{F-th13-chisq-SNO2005-single.eps}
        \caption{\label{fig:alpha}%
      Determination of $\alpha \equiv \Dms / \Dma$ and bound on
      $\sin^2\theta_{13}$ from current neutrino oscillation data.}
\end{figure}

The left panel in Fig.~\ref{fig:alpha} gives the parameter $\alpha$,
namely the ratio of solar over atmospheric splittings, as determined
from the global $\chi^2$ analysis of \cite{Maltoni:2004ei}.

\begin{figure}[t] \centering
    \includegraphics[height=4cm,width=.75\linewidth]{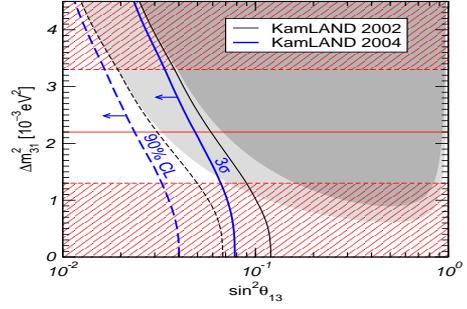}
    \caption{\label{fig:t13-solar-chooz} Upper bound on
      $\sin^2\theta_{13}$ (1 \dof) from solar and reactor data versus
      $\Dma$. Dashed (solid) curves correspond to 90\% (3$\sigma$)
      \CL\ bounds, thick curves include KamLAND-2004 data, thin ones
      do not.  Light (dark) regions are excluded by CHOOZ at 90\%
      (3$\sigma$) \CL\ The horizontal line corresponds to the current
      $\Dma$ best fit value, hatched regions are excluded by
      atmospheric + K2K data at 3$\sigma$.}
\end{figure}

The right panel in Fig.~\ref{fig:alpha} gives $\Delta\chi^2$ as a
function of $\sin^2\theta_{13}$ for different data samples.  One finds
that the KamLAND-2004 data have a surprisingly strong impact on this
bound. Before KamLAND-2004 the bound on $\sin^2\theta_{13}$ from
global data was dominated by the CHOOZ reactor experiment, together
with the determination of $\Delta m^2_{31}$ from atmospheric data.
However, including KamLAND-2004 the bound becomes comparable to the
reactor bound. Note also that, since the reactor bound on
$\sin^2\theta_{13}$ deteriorates quickly as $\Dma$ decreases (see
Fig.~\ref{fig:t13-solar-chooz}), the improvement is especially
important for lower $\Dma$ values.
In Fig.~\ref{fig:t13-solar-chooz} we show the upper bound on
$\sin^2\theta_{13}$ as a function of $\Dma$ from CHOOZ data alone
compared to the bound from an analysis including solar and reactor
neutrino data. One sees that, although for larger $\Dma$ values the
bound on $\sin^2\theta_{13}$ is dominated by CHOOZ, for $\Dma \lsim 2
\times 10^{-3} \eVq$ the solar and KamLAND data become relevant.

Altogether, the bound on $\sin^2\theta_{13}$ contributes significantly
to the overall global bound 0.047 at 3$\sigma$ for 1 \dof\ As shown
in~\cite{Maltoni:2004ei} such an improved $\sin^2\theta_{13}$ bound
follows mainly from the strong spectral distortion found in the 2004
sample.

Future long baseline reactor and accelerator neutrino oscillation
searches~\cite{Lindner:2005af}, as well as studies of the day/night
effect in large water Cerenkov solar neutrino experiments such as UNO
or Hyper-K~\cite{SKatm04} could bring more information on
$\sin^2\theta_{13}$~\cite{Akhmedov:2004rq}. With neutrino physics
entering the precision age it is necessary to scrutize also the
validity of the unitary approximation of the lepton mixing matrix in
future experiments, given its theoretical
fragility~\cite{schechter:1980gr}.
                                                                           
\section{Absolute neutrino mass scale}

\begin{figure}[b]
  \centering
\includegraphics[width=4.5cm,height=2.2cm]{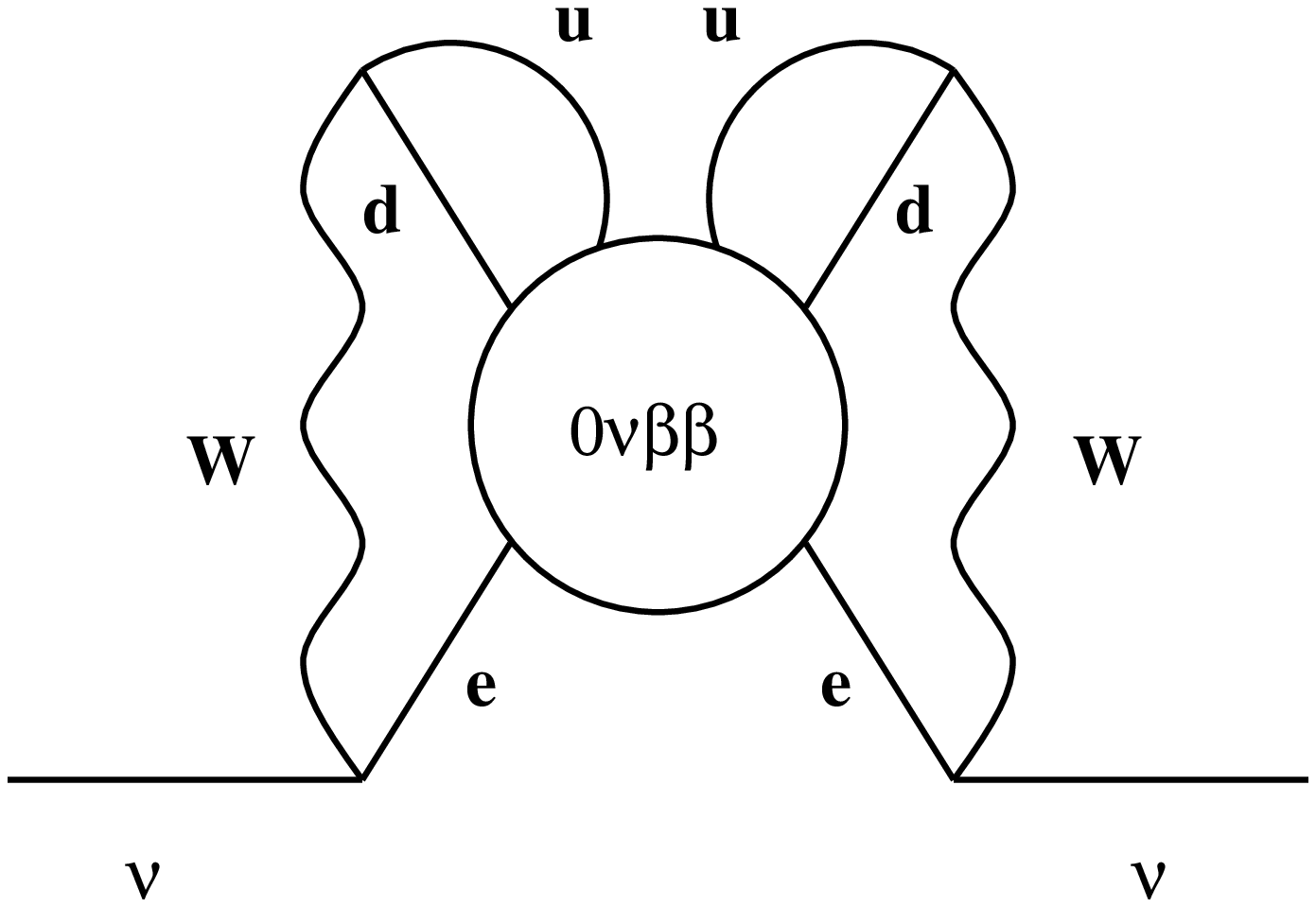}
  \caption{\nbb and Majorana mass are equivalent~\cite{Schechter:1981bd}.}
 \label{fig:bbox}
\end{figure}

On general grounds neutrino masses are expected to be
Majorana~\cite{schechter:1980gr}, a fact that may explain their
relative smallness with respect to other fermion masses. 
Neutrino oscillation data are insensitive to the absolute scale of
neutrino masses and also to the fundamental issue of whether neutrinos
are Dirac or Majorana particles~\cite{schechter:1981gk,doi:1981yb}.
Hence the importance of neutrinoless double beta
decay~\cite{Wolfenstein:1981rk}. The significance of the \nbb decay is
given by the fact that, in a gauge theory, irrespective of the
mechanism that induces \nbb, it is bound to also yield a Majorana
neutrino mass~\cite{Schechter:1981bd}, as illustrated in Fig.
\ref{fig:bbox}.
Quantitative implications of the ``black-box'' argument are
model-dependent, but the theorem itself holds in any ``natural'' gauge
theory.

Now that oscillations are experimentally confirmed we know that \nbb
must be induced by the exchange of light Majorana neutrinos. The
corresponding amplitude is sensitive both to the absolute scale of
neutrino mass as well as the two Majorana CP phases that characterize
the minimal 3-neutrino mixing matrix~\cite{schechter:1980gr}.
\begin{figure}[t]
  \centering
\includegraphics[width=.7\linewidth,height=5.5cm]{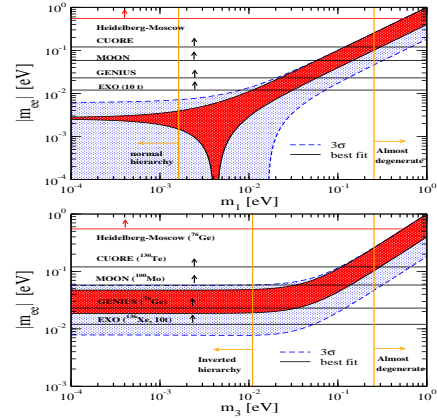}
  \caption{\nbb amplitude and current oscillation data.}
 \label{fig:nbbfut}
\end{figure}
Fig. \ref{fig:nbbfut} shows the estimated average mass parameter
characterizing the neutrino exchange contribution to \nbb versus the
lightest neutrino mass.  The calculation takes into account the
current neutrino oscillation parameters from \cite{Maltoni:2004ei} and
the nuclear matrix elements of~\cite{Bilenky:2004wn} and compares with
experimental sensitivities.
The upper (lower) panel corresponds to the cases of normal (inverted)
neutrino mass spectra. In these plots the ``diagonals'' correspond to
the case of quasi-degenerate
neutrinos~\cite{caldwell:1993kn,babu:2002dz}, which give the largest
\nbb amplitude.
In contrast to the normal hierarchy, where a destructive interference
of neutrino amplitudes is possible, the inverted neutrino mass
hierarchy implies a ``lower'' bound for the \nbb amplitude.
An exception to the rule that there is no lower bound on \nbb in
normal hierarchy models is provided by the model in
\cite{Hirsch:2005mc}.

Future experiments~\cite{Aalseth:2002rf} will extend the sensitivity
and provide an independent check of the Heidelberg-Moscow
claim~\cite{Klapdor-Kleingrothaus:2004wj}.
More information on the absolute scale of neutrino mass will also come
from future beta decays searches~\cite{Osipowicz:2001sq} and
cosmology~\cite{Hannestad:2004nb}.
            
\section{The origin of neutrino mass}

Neutrino mass arise from the dimension-five operator $\ell \ell \phi
\phi$ where $\phi$ the \21 Higgs doublet and $\ell$ is a lepton
doublet~\cite{Weinberg:1980bf}.  Nothing is known from first
principles about the mechanism that induces this operator, its
associated mass scale or flavour structure.  The neutrino masses that
result from it once the electroweak symmetry breaks down are expected
to be Majorana. This may explain why neutrino masses are much smaller
than those of the other fermions. This may happen either because the
operator is suppressed by a large scale in the denominator, or else
suppressed by a small scale in the numerator.  Both ways are viable
and can be made natural.

The most popular case is the seesaw mechanism which induces small
neutrino masses from the exchange of heavy states that may come from
unification. 
\begin{figure}[t] \centering
    \includegraphics[height=2cm,width=.45\linewidth]{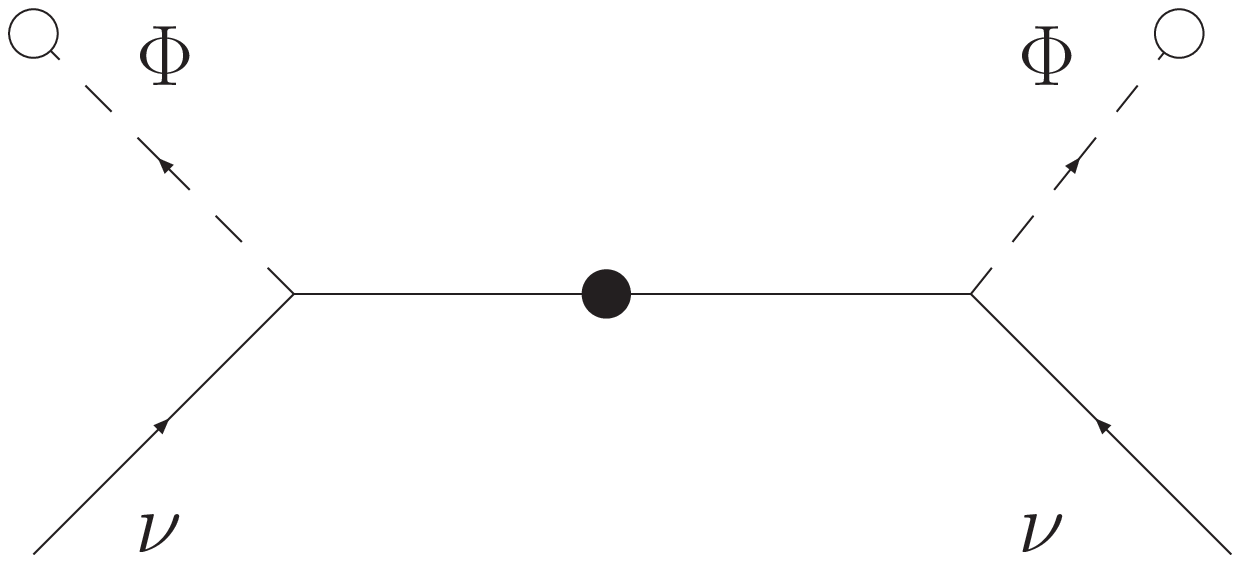}
    \includegraphics[height=2.2cm,width=.45\linewidth]{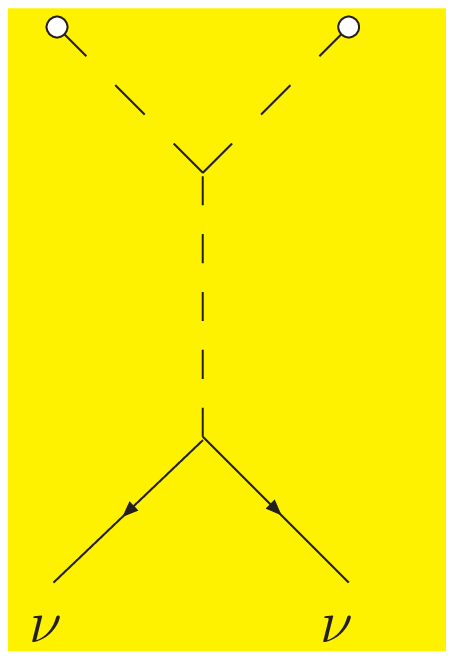}
    \caption{\label{fig:seesaw} %
     The seesaw mechanism.}
\end{figure}
Small neutrino masses are induced either by heavy \21 singlet
``right-handed'' neutrino exchange (type I) or heavy scalar bosons
exchange (type II), in a nomenclature opposite from the original one
in~\cite{schechter:1980gr}. The effective triplet seesaw term has a
flavor structure different from the type-I term, contributing to the
lack of predictivity of seesaw schemes. An attempt to recover
predicitvity within the seesaw approach by appealling to extra
symmetries~\cite{mnuth} is given in~\cite{babu:2002dz}.
The model predicts maximal $\theta_{23}$, $\theta_{13}=0$, and
naturally large $\theta_{12}$, though unpredicted. Moreover, if CP is
violated $\theta_{13}$ becomes arbitrary but the Dirac CP violation
phase is maximal~\cite{Grimus:2003yn}. The model gives a lower bound
on the absolute neutrino massd $m_{\nu}\gsim 0.3$ eV, requires a light
slepton below 200 GeV, and gives large rates for flavour violating
processes.

Amongst ``bottom-up'' models we mention those where neutrino masses
are given as radiative corrections~\cite{zee:1980ai,babu:1988ki} and
models where low energy supersymmetry is the origin of neutrino
mass~\cite{Hirsch:2004he}.  The latter are based on the idea that R
parity spontaneously break~\cite{Masiero:1990uj}, leading to a very
simple effective bilinear R parity violation model~\cite{Diaz:1997xc}.
In this case the neutrino spectrum is typically hierarchical, with the
atmospheric scale generated at the tree level and the solar scale
radiatively ``calculable''~\cite{Hirsch:2000ef}. For the parameters
that reproduce the masses indicated by current data, typically the
loghtest supersymmetric particle decays in the detector, and its decay
properties correlate with neutrino mixing angles, a test that can be
made, e.~g. at the LHC.

\section{Robustness of oscillations}

\begin{figure}[t] \centering
    \includegraphics[height=2cm,width=.6\linewidth]{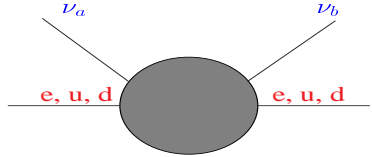}
    \caption{\label{fig:nuNSI} %
      Flavour-changing effective operator.}
\end{figure}

The general effective model-independent description of the seesaw at
low-energies was given in \cite{schechter:1980gr}. It is characterized
by (n,m), n the number of \21 isodoublet and m the number of \21
isosinglet leptons.
In the mass basis a the (3,3) seesaw model has 12 mixing angles and 12
CP phases (both Dirac and Majorana-type) characterizing its full
3$\times$6 charged current seesaw lepton mixing matrix and
non--diagonal neutral current~\cite{schechter:1980gr}.
The nontrivial structure of charged and neutral current weak
interactions with non-unitary lepton mixing matrix is a general
feature of seesaw models~\cite{schechter:1980gr} and lead to
dimension-6 terms non-standard neutrino interactions (NSI), as
illustrated in Fig.~\ref{fig:nuNSI}.
Such sub-weak strength $\varepsilon G_F$ operators can be of two
types: flavour-changing (FC) and non-universal (NU). 
In inverse seesaw-type models~\cite{mohapatra:1986bd} the non-unitary
piece of the lepton mixing matrix can be sizable and hence the induced
NSI may be phenomenologically important~\cite{bernabeu:1987gr}.
Sizable NSI strengths may also arise in radiative neutrino mass models
and in supersymmetric unified models~\cite{hall:1986dx}.
                                                                
Non-standard physics may in principle affect neutrino propagation
properties and detection cross sections~\cite{pakvasa:2003zv}.
In their presence, the Hamiltonian describing atmospheric neutrino
propagation has, in addition to the standard oscillation part, another
term $H_\mathrm{NSI}$ 
\begin{equation}
    H_\mathrm{NSI} = \pm \sqrt{2} G_F N_f
    \left( \begin{array}{cc}
        0 & \varepsilon \\ \varepsilon & \varepsilon'
    \end{array}\right) \,.
\end{equation}
Here $+(-)$ holds for neutrinos (anti-neutrinos) and $\varepsilon$ and
$\varepsilon'$ parameterize the NSI: $\sqrt{2} G_F N_f \varepsilon$ is
the forward scattering amplitude for the FC process $\nu_\mu + f \to
\nu_\tau + f$ and $\sqrt{2} G_F N_f \varepsilon'$ represents the
difference between $\nu_\mu + f$ and $\nu_\tau + f$ elastic forward
scattering. Here $N_f$ is the number density of the fermion $f$ along
the neutrino path.  In the 2--neutrino approximation, the
determination of atmospheric neutrino parameters $\Dma$ and
$\sin^2\theta_\Atm$ was shown to be practically unaffected by the
presence of NSI on down-type quarks ($f=d$)~\cite{fornengo:2001pm}.
Future neutrino factories will substantially improve this
bound~\cite{huber:2001zw}. 

In contrast, the oscillation interpretation of solar neutrino data is
``fragile'' in the presence of non-standard
interactions~\cite{Miranda:2004nb}, with a new ``dark side'' solution
(with $\sin^2\theta_{sol}\simeq 0.7$~\cite{Miranda:2004nb}),
essentially degenerate with the conventional one, present even after
combining with data from reactors.
On the other hand, it has been shown~\cite{huber:2002bi} that, even a
small residual non-standard interaction of neutrinos in the $e-\tau$
channel leads to a drastic loss in sensitivity in the $\theta_{13}$
determination at a neutrino factory. It is therefore important to
improve the sensitivities on NSI, another window of oportunity for
neutrino physics in the precision age.
                                                                         
Work supported by Spanish grant BFM2002-00345 and by the EC RTN
network MRTN-CT-2004-503369.


\end{document}